\documentclass[epj]{svjour}
\usepackage{graphics}
\usepackage{amsfonts}
\usepackage{amssymb}
\usepackage{amsmath}
\usepackage{graphicx}
\usepackage{float}
\usepackage{caption}
\usepackage{multicol}
\usepackage{dcolumn}

\begin{document}

\title{Primordial black holes and gravitational waves in teleparallel Gravity%
}
\author{K. El Bourakadi\inst{1}\thanks{%
k.elbourakadi@yahoo.com}, B. Asfour\inst{2}\thanks{%
brahim.asfour95@gmail.com}, Z. Sakhi\inst{1}\thanks{%
zb.sakhi@gmail.com}, Z. M. Bennai\inst{1}\thanks{%
mdbennai@yahoo.fr} \and Taoufik Ouali\inst{2}\thanks{%
t.ouali@ump.ac.ma}}
\offprints{}
\institute{Quantum Physics and Magnetism Team, LPMC, Faculty of Science Ben M'sik, 
{\small Casablanca Hassan II University, Morocco \and Laboratory of Physics
of Matter and Radiations, Mohammed first University, BP 717, Oujda, Morocco.}}
\date{Received: date / Revised version: date}

%

%

%
\abstract{
In this paper, we consider the possible effect of the teleparallel gravity
on the production of the primordial black holes (PBH) and on the
gravitational waves (GWs). We investigate the relationship between the slow
roll, the e-folds number and the teleparallel parameters. We show that in
the case of the teleparallel parameter $\delta =3$, the e-folds number
reaches the values $50$ and $60$ in consistency with the contour plot of the 
$(r,n_{s})$ plane obtained by Planck data at $1\sigma $ and $2\sigma $ C.L..
Furthermore, we use the fraction of the energy density and the variance of
the density perturbations approach to study the abundance of the production
PBH. We find that the PBH overproduction can be satisfied for specific
values of parameters of the non-adiabatic curvature power spectrum at some
narrow parametric resonance. Moreover, to explain the GWs expected by
observations, the duration of preheating should be bounded by values less
or equal to $2$. This bound is in agreement with values of the
tensor-to-scalar ratio and the spectral index constrained by Planck data.
\PACS{
      {PACS-key}{discribing text of that key}   \and
      {PACS-key}{discribing text of that key}
     } } 
\maketitle
\section{Introduction}

\label{intro} General theory of relativity (GR) successfully describes many
observational phenomena either at astrophysical \cite%
{will-2014,ligo-BH2016,ligo-GW2016,ligo2019} and cosmological \cite%
{SN-1999,SN-1998,WMAP-2003,Scolnic-2018,planck-2020} scales. However, this
theory faces many issues such as the standard big bang problems explained in
the context of the inflationary paradigm \cite{Guth-1918,Linde-1982}.
Indeed, among the most exciting theories that explain the origins of the
Universe is the inflation scenario \cite{A1,A2,A3}. This paradigm, described
through various models, gives a satisfying solution to the standard big bang
problems. Accurate measurements of the cosmic microwave background (CMB) and
the large-scale structure (LSS) of galaxies provide helpful constraints on
this inflationary paradigm \cite{A4}. Furthermore, at late-time, General
theory of relativity cannot explain the current speed of the expansion of
the Universe \cite{Weinberg-1989} unless an unknown exotic energy density
dubbed Dark energy is introduced. Other observations that form a stumbling
block for general relativity may be find in \cite%
{Malquarti-2003,Verde-2019,Riess-2021}. To overcome these issues, many
alternatives to GR have been suggested \cite%
{Clifton-2012,Kase-2019,Kobayashi-2019,Bahamonde-2021,CANTATA-2021}. {An
interesting alternatives that has gained more attention recently are $f(R)$
gravity \cite{Felice2010}, $f(G)$ gravity \cite{Nojiri2005} and $f(T)$
teleparallel gravity \cite{Hayashi-1967,Hayashi1982,Aldrovandi2013,Maluf2013}%
.} In this formalism, the geometrical deformation that produces
gravitational field is originated by the scalar torsion and not by the
scalar curvature. Teleparallel gravity is one of the alternative theories of
gravity that gives successful descriptions of the late-time acceleration as
well as the inflationary expansion phase {\cite{I1}}.{\ Even though
teleparallel gravity reproduces GR in its classical level, many developments
of such modified theories has received a lot of attraction \cite%
{Li2011,Bamba2012,Wu-2012,Ong2013,Otalora2013,Bahamonde2015,Rezazadeh2016,Farrugia2016,Awad2017}%
. To discriminate among the number of modified theories of gravity, further
analysis should be done in order to discard some of those theories which
disagree with observations. }This alternative theory of GR were considered
in many contexts such as in linear perturbations \cite%
{Wu-2012,Rezazadeh-2017}, inflation and reheating process \cite{B3}.\newline

The abundance of primordial black holes (PBH) depends on the primordial
power spectrum obtained at the end of inflation{\ and could be considered as
a candidate for the formation of dark matter \cite%
{Car2016,Gaggero2017,Inomata2017,Kovetz2017,Georg2017}.} Indeed, PBH forms
due to the amplification in the primordial power spectrum on small scales 
\cite{C1}. Assuming that the Universe has reached the adiabatic limit when
it is radiation-dominated, the comoving curvature perturbation is frozen on
superhorizon scales. When the scale re-enters the Hubble horizon, some
regions may have a large positive curvature which is equivalent to a closed
Universe. At this step the expansion eventually stops and the contraction
starts, then the Hubble size region with large positive curvature will
collapse to a black hole \cite{C1}. Furthermore, it has been argued that
after inflation, the inflaton condensate energy density may collapses and
forms black holes \cite{C2}. After inflation, the Universe enters a
reheating era preceded by an era called preheating. In the preheating phase
inflaton field began to oscillate and decay into massive bosons due to the
parametric resonance. In our model we focus on the preheating period which
occurs via a broad parametric resonance. We study the formation of PBH
during these period knowing that the scale that exit outside the Hubble
radius towards the end of inflation could possibly re-enter the Hubble
radius after reheating during the radiation dominated era. However,
amplified fluctuations during preheating lead to PBH formation on the
slightly shorter scales which re-enter the Hubble radius during preheating 
\cite{C3}. Moreover, these amplifications during the preheating process can
lead to the amplification of sufficiently large curvature perturbations and
lead to the overproduction of primordial black holes \cite{C3}. {The
imprints that PBH may leave in observations provide an important background
to some astrophysical issues. Indeed, these signatures may interpret the non
linear seeds of the large scale structure and the PBH evaporation could
explain the point-like gamma-ray sources \cite{Belotsky2014,Khlopov2010}.}%
\newline

On the other hand, gravitational waves, which is theoretically predicted by
general relativity, is originated from astronomical objects or by the
dynamical expansion of the early Universe. While merging black hole or
colliding neutron stars may describe the first generation, the so called
primordial GWs are generated in the early Universe. In this paper, {we
consider the second generation of GWs which is generated during the
inflationary era by means of the enhancing of the primordial curvature
perturbations. However, a significant production of GWs is sourced in the
preheating era characterized by the parametric resonance \cite{3,Khlebnikov}%
. These GWs leave an indirect imprint in the cosmic microwave background
temperature anisotropies, specially in the measurement of its polarization 
\cite{6}. } At the end of inflation, inhomogeneities of the time-dependent
field act as a gravitational source and the spectrum of GWs can be linked
directly to the duration of preheating. Inflationary scenario predicts such
GWs and their detection, whether directly or indirectly, can provide us a
unique opportunity to test theories of inflation \cite{2}. The amplitude of
the GWs spectrum, which is generated during preheating, is independent of
the energy scale of inflation characterizing the present peak frequency of
the GWs spectrum \cite{4,5}. \newline

Motivated by the increasing interest in teleparallel gravity, we study the
production of the primordial black hole, the constraints of the amplitude of
the scalar primordial on the parameters of the model. The gravitational
waves are also considered in this context. To this aim, we introduce the
inflationary paradigm and relate its parameters to the subsequent era namely
the preheating and the reheating phases. \newline

The paper is organized as follows: In Sec. \ref{secII}, we setup the basic
concept of teleparallel gravity and introduce some fundamental parameters of
the inflationary scenario. In Sec. \ref{secIII}, we constrain the e-fold
parameter of the preheating era to those of inflationary paradigm. In Sec. %
\ref{secIII-1}, we derive the reheating phase in terms of the e-folds number 
$N_{re}$, we brievely discuss the perturbative proccess of reheating and
highlight the correlation of $f(T)$ model parameters and the reheating
duration. In Sec. \ref{secIV}, we introduce the primordial black holes and
their production by means of model parameters. In Sec. \ref{secV}, we study
the production of the gravitational waves and their constrains from
observations. Sec. \ref{secVI} is dedicated to conclusions.

\section{Teleparallel gravity and inflationary parameters}

\label{secII}

As an alternative formulation of gravity, teleparallel gravity is a theory
formulated in terms of torsion with no curvature term \cite{B7}. The
fundamental orthonormal field which maps space-time coordinates $x^{\mu }$\
to the coordinates $x^{a}$\ on the tangent-space is the tetrad $e^{a}~_{\mu }
$. In terms of the tetrad, the space-time metric can be expressed as follows 
\cite{B7-1} 
\begin{equation*}
g_{\mu \nu }=\eta _{ab}e_{\;\;\mu }^{a}e_{\;\;\nu }^{b},
\end{equation*}%
where $\eta _{ab}$ is the Minkowski metric. The tetrad obeys normalization
conditions $e_{\;\;\mu }^{a}e_{b}^{\;\;\mu }=\delta _{b}^{a}$ and $%
e_{\;\;\mu }^{a}e_{a}^{\;\;\nu }=\delta _{\mu }^{\nu }$. The Latin and Greek
indices stand for tangent space and space-time coordinates, respectively.
Modified gravity can also be constructed by a given action based on
teleparallel gravity. We construct the modified teleparallel gravity by an
action depending on the torsion scalar $T,$ minimally coupled to a
barotropic fluid described by the scalar field, $\phi $, and the radiation's
Lagrangian density, $\pounds _{r}$. The action under consideration is then
given by \cite{B1} 
\begin{equation}
S=\int d^{4}x\sqrt{-g}\left[ \frac{M_{p}^{2}}{2}f(T)+\frac{1}{2}\partial
_{\mu }\phi \partial ^{\mu }\phi -V(\phi )+\pounds _{r}\right] ,  \label{S}
\end{equation}%
where $g$ is the determinant of the metric. The torsion scalar $T$ is
constructed by a contraction of the torsion tensor 
\begin{equation}
T=\frac{1}{4}T_{\rho \mu \nu }T^{\rho \mu \nu }+\frac{1}{2}T^{\nu \mu \rho
}T_{\rho \mu \nu }-T_{\rho \mu }^{\;\;\;\rho }T_{\;\;\;\;\nu }^{\nu \mu }
\end{equation}%
with the torsion tensor given by 
\begin{equation}
T_{\mu \nu }^{\gamma }=\Gamma _{\mu \nu }^{\gamma }-\Gamma _{\nu \mu
}^{\gamma }=e_{a}^{\;\;\gamma }\left( \partial _{\mu }e_{\;\;\nu
}^{a}-\partial _{\nu }e_{\;\;\mu }^{a}\right) .  \label{Tor}
\end{equation}%
In the context of the flat Friedmann-Lema\^{\i}tre-Robertson-Walker (FLRW)
metric 
\begin{equation}
ds^{2}=-dt^{2}+a^{2}(t)\left( dr^{2}+r^{2}(d\theta ^{2}+\sin ^{2}\phi
^{2}\right) ,
\end{equation}%
the tetrad field of a FLRW Universe is given by
\begin{equation}
e_{\;\;\mu }^{a}=\text{diag}(1,a(t),a(t),a(t)),
\end{equation}%
and the modified Friedman equations in teleparallel gravity are

\begin{eqnarray}
H^{2} &=&\frac{1}{3M_{p}^{2}}\left( \rho _{\phi }+\rho _{r}+\rho _{T}\right)
, \\
\dot{H} &=&-\frac{1}{2M_{p}^{2}}\left( \rho _{\phi }+\rho _{r}+\rho
_{T}+P_{\phi }+P_{r}+P_{T}\right) ,
\end{eqnarray}%
where $H$ and $\dot{H}$\ are the Hubble parameter and its differentiation
with respect to the cosmic time $t$. The energy density of the scalar field
and its pressure are given by $\rho _{\phi }=\dot{\phi}^{2}/2+V(\phi )$ and $%
P_{\phi }=\dot{\phi}^{2}/2-V(\phi )$, respectively. $\rho _{r}$ and $P_{r}$
are the energy density and the pressure of radiation. In the FLRW, the
torsion scalar is related to the Hubble parameter as $T=-6H^{2}$, the energy
density $\rho _{T}$\ and the pressure $P_{T}$ are writte as \cite{B1}

\begin{eqnarray}
\rho _{T} &=&\frac{M_{p}^{2}}{2}\left( 2Tf_{,T}-f-T\right) , \\
P_{T} &=&-\frac{M_{p}^{2}}{2}(-8\dot{H}Tf_{,TT}+\left( 2T-4\dot{H}\right)
f_{,T}-f  \notag \\
&&+4\dot{H}-T),
\end{eqnarray}%
where 
\begin{equation}
f_{,T}=\frac{\partial f(T)}{\partial T}\quad \text{and}\quad f_{,TT}(T)=%
\frac{\partial ^{2}f}{\partial T^{2}}.
\end{equation}

In order to analyze preheating parameters, we introduce the model used in 
\cite{B3} to derive the inflationary parameters that are directly linked to
inflation parameters. Usually preheating is characterized by the production
of $\chi $-particles through the parametric resonance which could be either
narrow or broad resonance depending on a specific condition on the
parameters of the well-known Mathieu equation that describe $\chi $-field
evolution. To illustrate our purpose, we consider a power-law modified
teleparallel \cite{B2}

\begin{equation}
f(T)=CT^{1+\delta },
\end{equation}%
where $\delta $ is a positive integer and $C=1/M^{2\delta }$ with $%
M=10^{-4}M_{p}$ is a mass dimension constant.\newline
For later uses, we give the background and the perturbative parameters in
the slow-roll approximation. The first and the second slow-roll parameters,
defined as $\epsilon _{1}=-\dot{H}/H^{2}$ and $\epsilon _{2}=\dot{\epsilon}%
_{1}/H\epsilon _{1}$, are given by \cite{B2}

\begin{eqnarray}
\epsilon _{1} &=&-\frac{\left( -M_{p}^{2}C\left( \delta +\frac{1}{2}\right)
\right) ^{\left( \frac{1}{1+\delta }\right) }}{\left( 1+\delta \right) }%
\frac{V^{\prime }(\phi )^{2}}{V(\phi )^{\left( \frac{\delta +2}{\delta +1}%
\right) }}, \\
\epsilon _{2} &=&2\epsilon _{1}\left( \left( 2+\delta \right) -2\left(
1+\delta \right) \left( \frac{V^{\prime \prime }(\phi )V(\phi )}{V^{\prime
}(\phi )^{2}}\right) \right) .
\end{eqnarray}%
The e-folds number of inflation considered\ from the time of the horizon
crossing to the inflation's end, labeled by the index "$k$" and "$end$"
respectively, is written as

\begin{eqnarray}
N_{I} &\simeq &-\int_{\phi _{k}}^{\phi _{end}}\frac{3H^{2}}{V^{\prime }(\phi
)}d\phi  \notag \\
&=&\frac{2^{\frac{-\delta }{1+\delta }}}{\left( M_{p}^{2}(-1)^{\delta
}C\left( 1+2\delta \right) \right) ^{\frac{1}{1+\delta }}}\int_{\phi
_{end}}^{\phi _{k}}\frac{V(\phi )^{\frac{1}{1+\delta }}}{V^{\prime }(\phi )}%
d\phi .  \label{Ni}
\end{eqnarray}%
The spectral index and the power spectrum are given by \cite{B2} 
\begin{eqnarray}
n_{s}-1 &\simeq &-2\epsilon _{1}-\epsilon _{2}, \\
{\mathcal{P}}_{s} &\simeq &\left( \frac{H_{k}^{2}}{8\pi
^{2}M_{p}^{2}c_{s}^{3}\epsilon _{1}}\right) _{c_{s}k=aH},  \label{Ps}
\end{eqnarray}%
where $c_{s}$ denotes the sound speed and the power spectrum is evaluated at
the horizon crossing i.e. at $c_{s}k=aH$. The sound speed for our suggested
model is

\begin{equation}
c_{s}^{2}=\frac{f_{,T}}{f_{,T}+12H^{2}f_{,TT}}=\frac{1}{1+2\delta }.
\end{equation}%
The relation between inflationary parameters and cosmological perturbations
provides a useful way to constrain preheating parameters for a chosen model
of inflation. For the case of chaotic potential $V=m^{2}\phi ^{2}/2$\ which
is recently argued that is the simplest and perhaps the most elegant model 
\cite{D10}, the solution to the well known Klein Gordon equation
corresponding to the inflaton field is given by $\phi =\Phi \cdot \sin mt$
with $m$ is the inflaton mass and $\Phi $\ is the initial amplitude of the $%
\phi $-field oscillations. The spectral index for the chaotic potential is
given by $n_{s}-1\simeq -4\epsilon _{1}$\ since $\epsilon _{2}=2\epsilon
_{1} $. Next, we compute preheating parameters, in particular, the
inflationary e-folds $N_{I}$, the Hubble parameter during inflation $H_{k}$
and the potential at the end of inflation $V_{e}$. The e-folds number during
inflation can be calculated as%
\begin{equation}
N_{I}=4\cdot \frac{(-1)^{\frac{2}{1+\delta }}}{\left( 1-n_{s}\right) \left(
\delta -1\right) }.
\end{equation}%
In order to calculate the Hubble parameter $H,$ one needs to compute the
power spectrum given by \cite{B3}%
\begin{equation}
{\mathcal{P}}_{s}\simeq \left( -\frac{\left( -M_{p}^{2}C\left( \frac{1}{2}%
+\delta \right) \right) ^{\frac{2\delta }{1+\delta }}\left( 1+\delta \right) 
}{12\pi ^{2}C^{2}M_{p}^{6}\sqrt{1+2\delta }}\frac{M_{p}^{2}}{\alpha ^{2}}%
\frac{V(\phi )^{\frac{3-\delta }{1+\delta }}}{V^{\prime }(\phi )^{2}}\right)
_{c_{s}k=aH},
\end{equation}%
which in the case of chaotic inflation\ writes 
\begin{equation}
{\mathcal{P}}_{s}\simeq -\frac{8}{3}\frac{m^{2}}{\pi ^{2}C^{2}M_{p}^{6}\sqrt{%
1+2\delta }}\frac{\left( -M_{p}^{2}C\left( \frac{1}{2}+\delta \right)
\right) ^{2}}{(1+\delta )(1-n_{s})^{2}},
\end{equation}%
knowing that $\epsilon _{1}=(1-n_{s})/4,$ Eq. (\ref{Ps}) can be rewritten as%
\begin{equation}
H^{2}_k\simeq 2\pi ^{2}M_{p}^{2}c_{s}^{3}{\mathcal{P}}_{s}\left(
1-n_{s}\right) .
\end{equation}%
Inflation ends when the slow-roll parameter $|\epsilon _{1}(\phi
_{end})|\approx 1$\ and the value of the field at the end of inflation writes

\begin{equation}
\phi _{end}^{2}=2^{\delta +2}\left( \frac{1}{1+\delta }\right)^{1+\delta
}\left(M_{p}^{2}Cm^{2\delta }\left( \frac{1}{2}+\delta \right) \right) .
\end{equation}%
An essential inflationary parameter is the tensor-to-scalar ratio which is
given, in teleparallel gravity, by \cite{B2}%
\begin{equation}
r=16c_{s}^{3}\epsilon _{1}.
\end{equation}%
The observable quantities $r$ and $n_{s}$\ can also be obtained in terms of $%
N_{I} $\ as%
\begin{eqnarray}
r &=&\frac{16}{N_{I}\left( \delta -1\right) \sqrt{1+2\delta }},  \notag \\
n_{s}-1 &=&-\frac{4}{N_{I}\left( \delta -1\right) }.
\end{eqnarray}

\begin{figure}[tbp]
\resizebox{0.75\textwidth}{!}{  \includegraphics{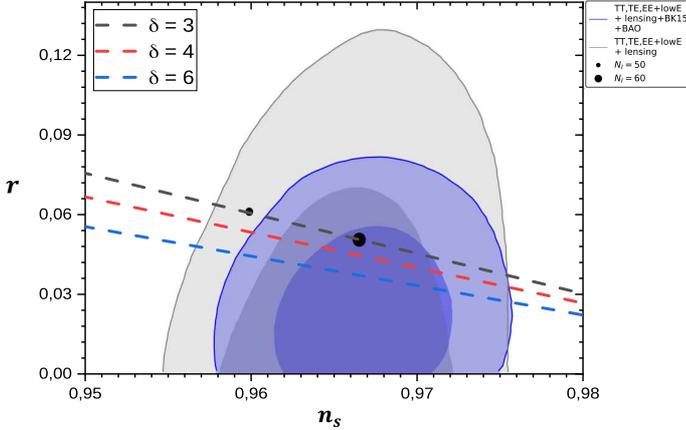}
} 
\caption{$r$ as a function of $n_{s}$ for a chaotic potential in modified
teleparallel gravity. Inner and outer shaded regions are $1\protect\sigma $
and $2\protect\sigma $ constraints in combination with CMB lensing
reconstruction and BAO from Planck data, respectively. We choose three
different values of $\protect\delta $, the black line for $\protect\delta =3$%
, the red line represents $\protect\delta =4$ and the blue line for $\protect%
\delta =6$. The dots correspond to $50$ and $60$ inflationary e-folds for
the case of $\protect\delta =3$, respectively.}
\label{fig:1}
\end{figure}

Considering the chaotic potential in our chosen teleparallel gravity, Fig. (%
\ref{fig:1}) shows that the tensor-to-scalar ratio, $r$, is a decreasing
function with respect to the spectral index, $n_{s}$. The results show a
good consistency for a specific range of model parameters with the latest
observations from Planck data. Moreover, in the case of $\delta =3$, the
choice of $50$\ and $60$ inflationary e-folds produces consistent
observational parameters with recent results. Finally, since $N_{I}$, $H$,
and $V_{e}$ are expressed as functions only of $\delta $ and $n_{s},$ one
may extend the investigation to the preheating parameter $N_{pre}$ and to
the so-called non-adiabatic curvature $\zeta _{nad}$ that\ arises from $\chi 
$ field perturbation. The next section will be devoted to this task.

\section{Constraints on preheating parameters}

\label{secIII}

In order to describe preheating through the parametric resonance, one would
introduce a preheating field $\chi .$ During preheating the potential gets
an additional term \cite{D12}%
\begin{equation}
V(\phi ,\chi )=m^{2}\phi ^{2}/2+g^{2}\phi ^{2}\chi ^{2}
\end{equation}%
with $g$ is the coupling of the inflaton to the $\chi $ field. After
inflation, amplified quantum fluctuations in $\chi $ field obey the wave
equation \cite{D11}%
\begin{equation}
\delta \ddot{\chi}+3H\delta \dot{\chi}+\left( \frac{k^{2}}{a^{2}}+g^{2}\phi
^{2}\right) \delta \chi =0,  \label{EoMx}
\end{equation}%
where $k$\ is the comoving wavenumber and $a$\ is the scale factor.\ The $%
\chi $ field with the effective mass results in an efficient preheating with
large amplitude oscillations when $q\equiv g^{2}\Phi ^{2}/4m^{2}\gg 1$. The
growth of the $\chi $ field fluctuations during preheating gives rise to the
non-adiabatic curvature perturbation $\zeta _{nad}$. Since the inflaton
decay can violate the adiabatic condition during preheating, pressure
perturbations can be split into adiabatic and non-adiabatic parts and the
evolution of $\zeta $ arises from the non-adiabatic part of pressure
perturbations. Moreover, the non-adiabatic perturbations can cause a change
in $\zeta $ on arbitrarily large scales when non-adiabatic pressure
perturbations are non-negligible. In fact, variations of $\zeta $ during
preheating could be driven by the non-adiabatic part of the $\chi $ field
perturbation. The power spectrum resulting from this amplification is given
by \cite{D11,D12}

\begin{equation}
{\mathcal{P}}_{\zeta _{nad}}\simeq \frac{2^{9/2}3}{\pi ^{5}\mu ^{2}}\left( 
\frac{\Phi }{M_{P}}\right) ^{2}\left( \frac{H_{end}}{m}\right) ^{4}\frac{%
g^{4}}{q^{\frac{1}{4}}}\left( \frac{k}{k_{end}}\right) ^{3}I(\kappa ,m\Delta
t),  \label{Pc}
\end{equation}%
where, by defining $\kappa ^{2}\equiv \frac{1}{18\sqrt{q}}\left( \frac{k}{%
k_{end}}\right) $, 
\begin{equation}
I(\kappa ,m\Delta t)\equiv \frac{3}{2}\int_{0}^{\kappa _{cut}}d\kappa
^{\prime }\int_{0}^{\pi }d\theta e^{2(\mu _{\kappa ^{\prime }}+\mu _{\kappa
-\kappa ^{\prime }})m\Delta t}\kappa ^{\prime 2}\sin (\theta ),  \label{0}
\end{equation}%
here $\theta $ is the angle between $\kappa ^{\prime }$ and $\kappa $, $%
\kappa _{cut}$ is an ultraviolet cut-off. The comoving wavenumber at the
Hubble radius exit and at the\ end of inflation are denoted by $k$\ and $%
k_{end}$, respectively. The term $m\Delta t$\ is an alternative way to
estimate how long the process of preheating will proceed and $\mu =(\ln
3)/2\pi $ for $q\gg 1$. At the end of inflation when $m\Delta t=0$, the
integral appeared in the power spectrum ${\mathcal{P}}_{\zeta _{nad}}$ is
estimated to $I(\kappa ,0)=\kappa _{cut}^{3}\sim 1$ \cite{D12}.

\begin{figure}[tbp]
\resizebox{0.85\textwidth}{!}{  \includegraphics{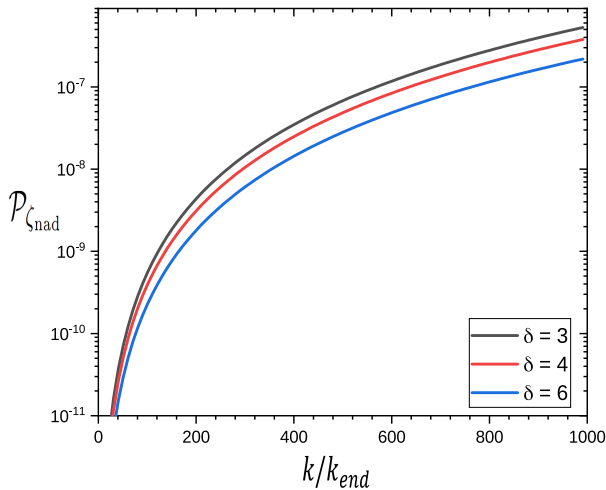}
} 
\caption{The non-adiabatic curvature perturbation ${\mathcal{P}}_{\protect%
\zeta _{nad}}$\ is presented for different $\protect\delta $\ values, $%
\protect\delta =3$, $4$\ and $6$. The other parameters are given as : $\Phi
\simeq M_{p}/\protect\sqrt{12\protect\pi },$ $g=10^{-3},$ $m=10^{-6}M_{p},$\ 
$q=6600$.}
\label{fig:2}
\end{figure}
From Eq.(\ref{Pc}), $H_{end}$\ is determined through the modified Friedmann
equation due to the teleparallel gravity under consideration and is given by 
\cite{B3}%
\begin{equation}
H_{end}^{2}=\frac{2^{\frac{-\delta }{\delta +1}}}{3}\left( \frac{V_{e}}{%
3\left( -1\right) ^{\delta }M_{p}^{2}C\left( 1+2\delta \right) }\right) ^{%
\frac{1}{1+\delta }}.
\end{equation}

The amplitude of non-adiabatic curvature perturbations\ ${\mathcal{P}}%
_{\zeta _{nad}}$\ is displayed as a function of $k/k_{end}$ in Fig. (\ref%
{fig:2}), for $\delta =3$, $4$, $6$, $m\Delta t=0$ and $I(\kappa ,0)=1.$\
One can check that when $k/k_{end}\rightarrow 0$,\ the power spectrum is in
the order of $\ \sim 10^{-11}$. On the other hand, when the ratio $k/k_{end}$
increases the power spectrum increases towards values higher than $\sim
10^{-7}$.\ Since ${\mathcal{P}}_{\zeta _{nad}}\propto k^{3}$, when $k\sim
k_{end}$\ and $m\Delta t\neq 0$ the change in $\zeta $ is weakly affected
for the large scale structure to be formed due to the preheating. In fact,
small scale fluctuations\ may become large enough for PBH to be overproduced 
\cite{D11}. We will discuss this in more detail in the next section.
Furthermore, extracting informations about preheating requires considering
the phase between the scale at the horizon crossing during inflation to the
present time, which can be derived in terms of inflationary parameters as 
\cite{B5,B6} 
\begin{equation}
N_{pre}=\left[ 61.6-\frac{1}{4}\ln \left( \frac{V_{e}}{\gamma H^{4}}\right)
-N_{I}\right] -\frac{1-3\omega }{4}N_{re},  \label{Npr}
\end{equation}%
where the parameter $\gamma $ is assumed to relate the energy density at the
end of inflation, $\rho _{e}$, to the preheating energy density, $\rho
_{pre} $. The reheating duration $N_{re}$ can be obtained from reheating
energy density $\rho _{re}=\left( \rho _{e}/\gamma \right) e^{-3\left(
1+\omega \right) N_{re}}$ \cite{B5}, with $\rho _{re}=\left( \pi
^{2}/30\right) g_{\ast }T_{re}^{4}$ and $g_{\ast }$\ denotes the number of
relativistic degrees of freedom at the end of reheating \cite{B5} 
\begin{equation}
N_{re}=\frac{1}{3(1+\omega )}\ln \left( \frac{3^{2}\cdot 5~V_{e}}{\gamma \pi
^{2}g_{\ast }T_{re}^{4}}\right) .  \label{Nr}
\end{equation}

\begin{figure*}[h]
\resizebox{1\textwidth}{!}{  \includegraphics{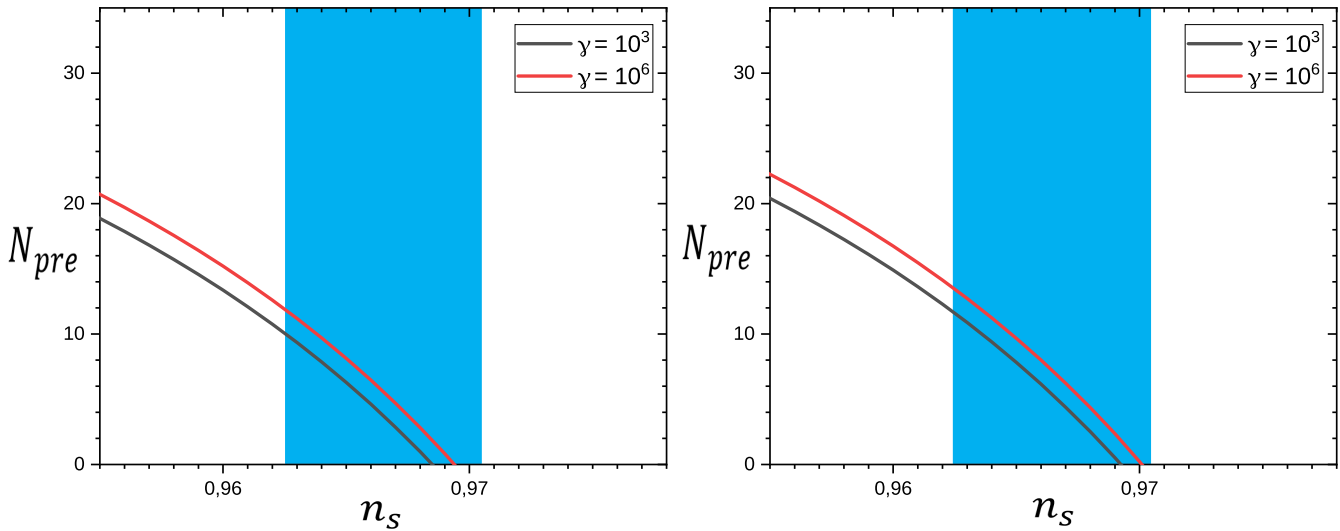}
}
\caption{The preheating $N_{pre}$ as functions of $n_{s}$ for chaotic
inflation. We choose $\protect\gamma $ to be $\protect\gamma =10^{3}$ and $%
\protect\gamma =10^{6}$. The blue region represents Planck's bound on $%
n_{s}. $ In the left panel, we considered the temperature $%
T_{re}=10^{12}GeV, $\ while in the right one, we choose a minimal reheating
temperature $T_{re}=100~GeV$.}
\label{fig:3}
\end{figure*}

We provide a numerical evaluation of the duration of preheating for the case
of chaotic potential in the modified teleparallel inflation as shown in Fig.
(\ref{fig:3}). We choose the EoS parameter to be $\omega \simeq 1/4$
following the assumption that during preheating the EoS gets closer to $1/3$ 
\cite{B5}, we take also a fixed value of $\delta =3$\ following constraints
on the inflationary e-folds found in the previous section. We observe that
the preheating duration is weakly sensitive to the reheating temperature and
is weakly shifted to the right for higher values of the parameter $\gamma $.
From Eq. (\ref{Npr}),\ lower values of reheating temperature allow a minimum
value of the preheating duration. In this direction, we choose two cases of
the final reheating temperature, the first one in order of $\sim 10^{12}GeV$
and the second one in order of the electroweak scale $\sim 100GeV$.\ Both
cases in the figure (\ref{fig:3}) show a good consistency with recent
observation results since all lines fall toward the central value of the
observational bound \cite{A4}. We notice that from Eq. (\ref{Nr}) the
maximum reheating temperature is bounded as $T_{re}\sim \lbrack
10^{12}-10^{13}]$ where reheating is defined to be instantaneous. However,
we consider a lower case where the reheating occurs at the electroweak scale
to test the thermalization temperature effects on the reheating duration. In
this model, preheating could appear either instantaneously or reach $%
N_{pre}\sim 10$\ to $14$\ e-folds depending on the choice of $T_{re}$ and $%
\gamma $ parameter. On the other hand, in our setup, preheating scenario is
very sensitive to the $\delta $\ parameter of the chosen teleparallel model $%
f(T).$ In fact, the main reason behind the choice of $\delta =3$\ is the
inflationary e-folds number $N_{I}$ that will take higher values when $%
\delta >3$\ which is inconsistant with Planck observation as shown in Fig. (%
\ref{fig:1}). Hence, higher values of $N_{I}$ affects the preheating
duration (see: Eq. (\ref{Npr})) and makes the curves fall away from the
central bound on $n_{s}.$ \ 

\section{Reheating in modified teleparallel gravity}

\label{secIII-1}

During preheating, the oscillating inflaton field decays into massive bosons 
$\chi $. However, it is preferred that there will be no explosive creation
of fermions taking into consideration the Pauli exclusion principle \cite%
{B5-1}. The explosive particles production during this step is due to the
parametric resonance when the amplitude of the inflaton field and the
coupling constants became large. From the measurements of CMB anisotropies
we know that the Universe reheated by reaching a thermal equilibrium in the
beginning of the big-bang nucleosynthesis $(BBN)$ with a temperature
satisfying $T_{re}>T_{BBN}$ \cite{B5-2}, this reheating phase occured from
the end of preheating at $a=a_{pre}$ and ended at $a=a_{re}$. Usually
reheating is discussed through pertubative approach \cite{B5-3,B5-4} which
describe the inflaton decay into relativistic bosonic and fermionic
particles. This decay is explained by inserting the friction term $\Gamma 
\dot{\phi}^{2}$ into the inflaton equation of motion, where $\Gamma $ is the
decay rate \cite{B5-5}.

During inflation, the scalar field decreases very slowly, and then after the
slow roll, a rapid oscillation starts at reheating phase, which generates
relativistic particles. The produced relativistic particles become dominant
and compose a state of thermal equilibrium fluid. During this rapid
oscillation until the radiation dominance, we can quantify the reheating
phase by the duration $N_{re}$ which takes into consideration the parameters
of our chosen modified teleparallel model of gravity. Knowing that energy
densities at the end of preheating and reheating respectively are given by $%
\rho _{re}\propto a_{re}^{-3(1+\omega )}$ and $\rho _{pre}=\rho
_{end}/\gamma =a_{pre}^{-3(1+\omega )}$\ \cite{B5}. The decay rate is given
in terms of the reheating temperature $T_{re}$\ and the number of
relativistic degrees of freedom $g_{\ast }$ \bigskip by \cite{B5-2}

\begin{equation}
\Gamma ^{2}\simeq \frac{\pi ^{2}}{10M_{p}^{2}}g_{\ast }T_{re}^{4}.
\end{equation}%
Hence, the duration of reheating is derived as

\begin{eqnarray}
N_{re} &=&\ln \left( \frac{a_{re}}{a_{pre}}\right) \\
&=&\frac{-1}{3(1+\omega )}\ln \left( \frac{\frac{2}{3}\gamma \Gamma ^{2}}{%
2^{\delta +2}\left( \frac{1}{1+\delta }\right) ^{1+\delta
}M_{p}^{2}Cm^{2+2\delta }\left( \frac{1}{2}+\delta \right) }\right) .  \notag
\end{eqnarray}%
\ \ 

This analysis about reheating is based on the original studies of
perturbative reheating after inflation where the ultra-relativistic
particles are gradually produced until the radiation era begins to dominate.
Our reheating duration result is compatible with the existence of a primary
preheating phase that is studied by non-perturbative methods. Here we
suppose that at this step the EoS is still moving towards $\omega =1/3$
where the radiation era takes place.

\section{Primordial black hole production}

\label{secIV}

Preheating is the period of the broad parametric resonance which occurs at
the end of inflation and before reheating. In addition, preheating is
characterized by amplified $\chi $ field fluctuations that are responsible
for a non-adiabatic curvature perturbation $\zeta _{nad}.$ The formation of
PBH in the early Universe is usually due to a collapse of large density
perturbations \cite{D3}. Some observational constraints on PBH abundance
have been considered in \cite{D13,D14} and were estimated to be less than $%
10^{-20}$ of the total energy density of the Universe. An important
condition that must be satisfied is that the density contrast must exceed
the critical value $\delta _{c}=0.414$ \cite{D8} in order for PBH to be
formed during the radiation domination era when fluctuations enter the
horizon. PBH production is estimated by means of the fraction of the energy
density which in turns is related to the variance of the density
perturbation. To this aim, we use the power spectrum and a window function $%
W(kR)$ \cite{C1}, to define the variance $\sigma (k)$ 
\begin{equation}
\sigma ^{2}(k)=\frac{16}{81}\int_{0}^{\infty }\left( \frac{\tilde{k}}{k}%
\right) ^{4}{\mathcal{P}}_{\zeta _{nad}}(\tilde{k})W^{2}(\tilde{k}R)\frac{d%
\tilde{k}}{\tilde{k}},
\end{equation}%
where the comoving horizon length $R=1/k$ is the scale at which the variance
is assumed to be coarse-grained. The window function has a Gaussian form
given by 
\begin{equation}
W(\tilde{k}R)=\exp \left( -\frac{\tilde{k}^{2}R^{2}}{2}\right) .
\end{equation}%
The final form of the variance of the density perturbations is obtained as 
\cite{C3} 
\begin{eqnarray}
\sigma ^{2}(k) &\approx &10\sqrt{2\pi }\frac{2^{9/2}3}{\pi ^{5}\mu ^{2}}%
\left( \frac{\Phi }{M_{p}}\right) ^{2}\left( \frac{H_{end}}{m}\right)
^{4}g^{4}q^{1/2}\left( \frac{k}{k_{end}}\right) ^{3}  \notag \\
&&I(\kappa ,m\Delta t).  \label{S(k)}
\end{eqnarray}

\begin{figure*}[h]
\resizebox{1\textwidth}{!}{  \includegraphics{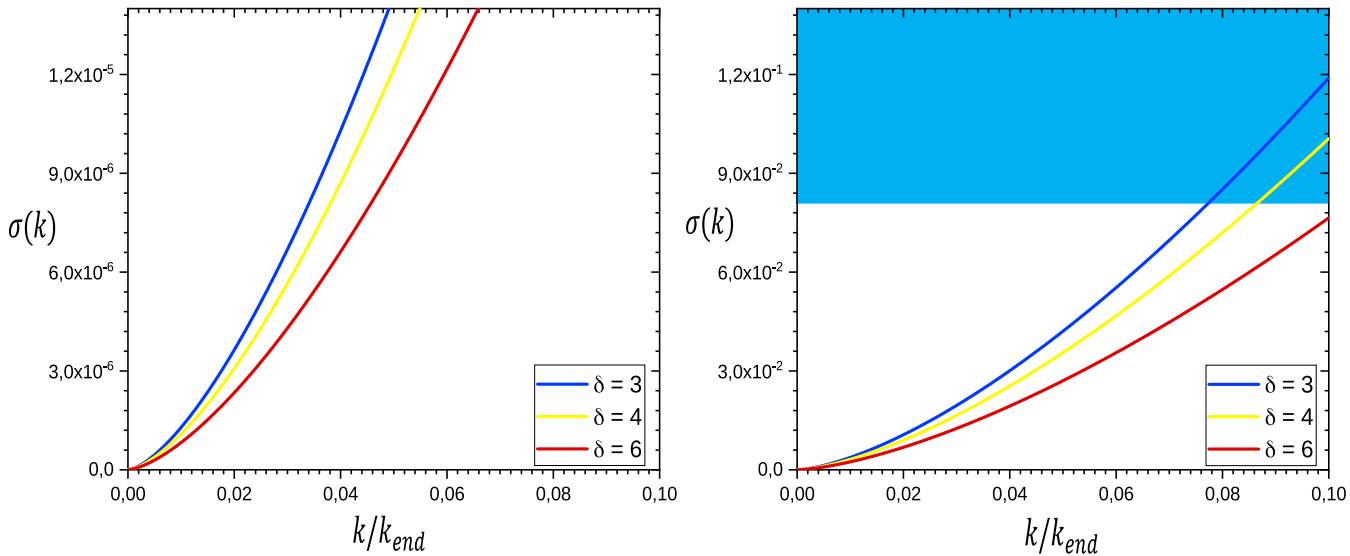}
}
\caption{The mass variance $\protect\sigma (k)$ as a function of
dimensionless wavenumber $k/k_{end}$, for $\protect\delta =3$ (blue-line), $%
\protect\delta =4$ (yellow-line), $\protect\delta =6$ (red-line), $\Phi
\simeq M_{p}/\protect\sqrt{12\protect\pi },g=10^{-3},m=10^{-6}M_{p}$ and $%
q=6600$. The blue region shows the threshold$\ \protect\sigma _{thresh}=0.08$
at which PBH are overproduced. The left panel presents the case of $m\Delta
t=25,$\ while in the right panel $m\Delta t=50$.}
\label{fig:4}
\end{figure*}

Evaluating Eq. (\ref{S(k)}),\ we plot the mass variance $\sigma (k)$ at the
horizon crossing\ in Fig. (\ref{fig:4}) as a function of the wavenumber
ratio $k/k_{end}$ for several values of $\delta $. According to \cite{D11}
at later times of preheating, the integral appearing in Eq. (\ref{0}) is
calculated as%
\begin{equation}
I(\kappa ,m\Delta t)=0.86(m\Delta t)^{-3/2}e^{4\mu m\Delta t}.
\end{equation}

We find that for $k/k_{end}\leq 10^{-1}$ and $m\Delta t=25,$ the variance
does not exceed the threshold\ value with $\sigma _{thresh}=0.08$\ \cite%
{D12,D4}\ at which the overproduction of PBH occurs. However, for the case $%
m\Delta t=50$ and $k/k_{end}\leq 10^{-1}$, the overproduction of PBH take
place for $\delta =3$, $4$ as $\sigma \geq \sigma _{thresh}.$

Assuming that primordial curvature perturbations follows Gaussian
distributions lead to estimating the abundance of PBH \cite{C1}. The
fraction of the energy density that collapses into PBH is estimated\ as \cite%
{D3}\ 
\begin{equation}
\beta (k)=\frac{1}{2}Erfc\left( \frac{\delta _{c}}{\sqrt{2}\sigma (k)}%
\right) .  \label{Beta}
\end{equation}%
The complementary error function can be approximated so that the energy
density fraction can be rewritten as \cite{D7,D8}%
\begin{equation}
\beta (k)\simeq \sqrt{\frac{1}{2\pi }}\frac{\sigma }{\delta _{c}}\exp \left(
-\left( \frac{\delta _{c}}{\sqrt{2}\sigma (k)}\right) ^{2}\right) .
\end{equation}

\begin{figure*}[h]
\resizebox{1\textwidth}{!}{  \includegraphics{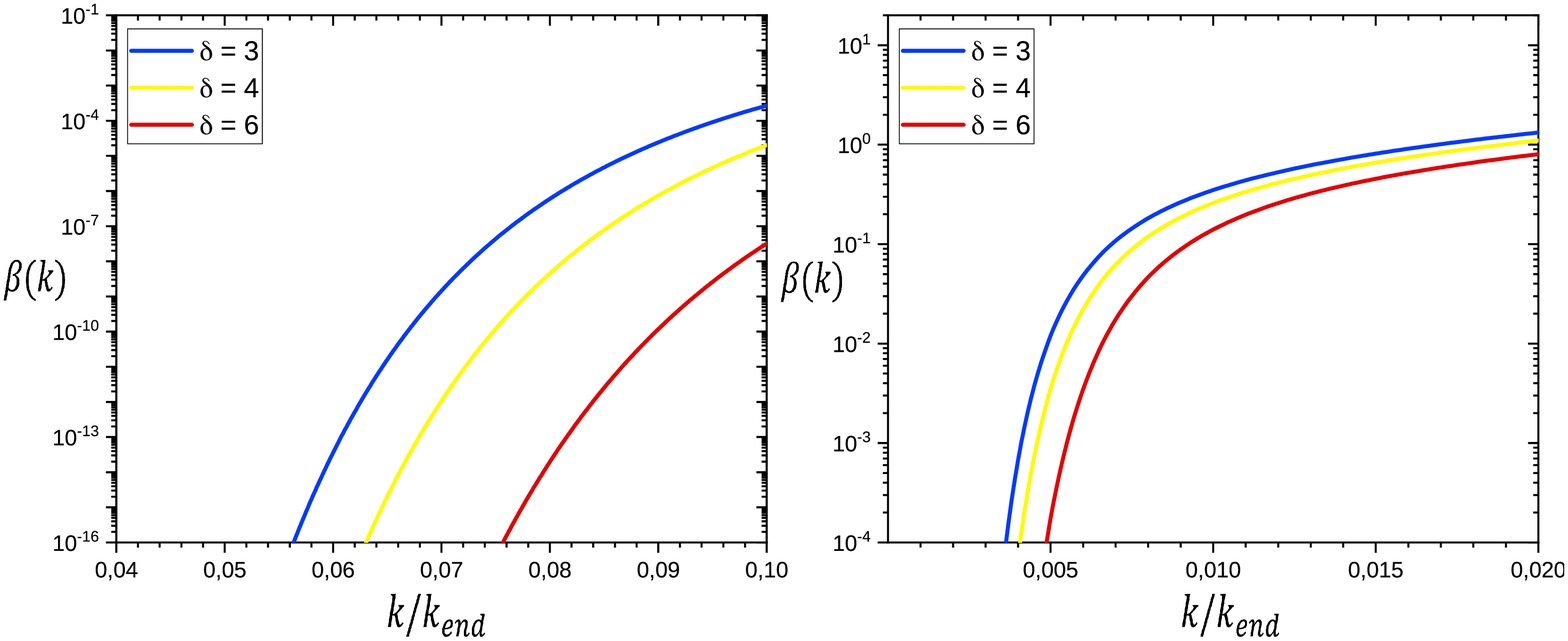}
}
\caption{The fraction of the total energy density collapsing into PBH as
functions of dimensionless wavenumber $k/k_{end}$ for $\protect\delta =3$
(blue-line), $\protect\delta =4$ (yellow-line) and $\protect\delta =6$
(red-line). The left panel presents the case of $m\Delta t=50,$\ while in
the right panel is for $m\Delta t=65$.}
\label{fig:5}
\end{figure*}

One important question is whether PBH are large enough to explain the LIGO-Virgo
events. For this reason, the fraction of the energy density can give an
alternative explanation to study PBH collapses in the early Universe. Fig. (%
\ref{fig:5}) shows the fraction of the total energy density for spherically
symmetric regions collapsing into PBH as a function of the dimensionless
wavenumber $k/k_{end}$. The results show that the curves converge toward $%
10^{-1}$\ for $m\Delta t=50$ where the overproduction of PBH is less
probable.\ However, for an overproduction regime where $m\Delta t=65$, PBH
production is satisfied for $\delta =3$, $4$ \ and $6$\ when $k/k_{end}\geq
0.015$.

\section{Gravitational Waves production}

\label{secV}

In this section we discuss Gravitational Waves production in $f(T)$\
gravity. The intense production of matter fields after inflation can promote
substantial metric changes. However, instead of the symmetric metric field,
we only focus on the tetrad field and the generation of GWs during
preheating. Since gravitational waves are detected through line element
change we no longer consider 10 components of the metric tensor, instead,
the 16 components of the tetrads must be taken into consideration and the
tetrad $e_{\mu }^{a}$ will have an additional tangent space-time indices 
\cite{C4,C5}. The metric satisfies the condition%
\begin{equation}
g_{\mu \nu }=\eta _{ab}e_{\;\;\mu }^{a}e_{\;\;\nu }^{b}=\eta _{ab}\bar{e}%
_{\;\;\mu }^{a}\bar{e}_{\;\;\nu }^{b},
\end{equation}%
and the tetrad can be decomposed as%
\begin{equation}
e_{\;\;\mu }^{a}=\bar{e}_{\;\;\mu }^{a}+\chi _{\;\;\mu }^{a},
\end{equation}%
where $\bar{e}_{\;\;\mu }^{a}$\ represents the part corresponding to metric
components, and $\chi _{\;\;\mu }^{a}$\ illustrates the degrees of freedom
obtained from the local Lorentz transformation. Focussing only on the $\bar{e%
}_{\mu }^{a}$ part, which can be perturbed around the flat FLRW background\
that gives rise to metric perturbations. We can obtain the perturbed torsion
tensor from these calculations that lead to perturbations in the field
equations from which the equation of motion for the GWs writes \cite{C6}%
\begin{equation}
\ddot{h}_{ij}+\left( 3H+\frac{\dot{f}_{,T}}{f_{,T}}\right) \dot{h}_{ij}-%
\frac{\nabla ^{2}}{a^{2}}h_{ij}=0
\end{equation}%
The energy density carried by GWs and sourced by the inhomogeneous decay of
the symmetry braking field is given by \cite{C7}

\begin{equation}
\rho _{GW}=\frac{M_{p}^{2}}{4}\langle \dot{h}_{ij}(t,x)\dot{h}%
_{ij}(t,x)\rangle
\end{equation}%
the abundance of gravity waves energy density today is presented by the
energy spectrum given as \cite{C7}

\begin{equation}
h^{2}\Omega _{GW,0}(f)=\frac{h^{2}}{\rho _{c,0}}\frac{d\rho _{GW,0}}{d\ln f},
\end{equation}%
where $h$ is the present dimensionless Hubble constant, $f$ is the frequency
and $\rho _{c,0}=3H_{0}^{2}/(8\pi G)$\ \ is the current critical energy
density. Considering the scale factor at the present time, $a_{0}$, and at
the time when GWs production stops, $a_{end}$, the GWs spectrum can be
converted into the actual physical quantities in order to correlate the
density spectrum with current observations \cite{C7} 
\begin{equation}
\frac{a_{end}}{a_{0}}=\frac{a_{end}}{a_{pre}}\left( \frac{a_{pre}}{a_{re}}%
\right) ^{1-\frac{3}{4}(1+\omega )}\left( \frac{g_{\ast }}{g_{0}}\right) ^{-%
\frac{1}{12}}\left( \frac{\rho _{r,0}}{\rho _{\ast }}\right) ^{\frac{1}{4}},
\label{GW0}
\end{equation}%
here we consider that at the end of preheating $"_{pre}"$ GWs production
stops. $"_{0}"$ and $"_{re}"$\ \bigskip indicate the present time and the
time when reheating is finished, respectively. $\rho _{\ast }$\ and $\rho
_{r,0}$\ are the total energy density of the scalar field and the current
radiation energy density, respectively. The present GWs spectra can finally
be given in the form \cite{B5}

\begin{equation}
\Omega _{GW,0}h^{2}=e^{-4N_{pre}}\left( \frac{g_{\ast }}{g_{0}}\right) ^{-%
\frac{1}{3}}\Omega _{r,0}h^{2}\Omega _{GW}(f),  \label{GW}
\end{equation}%
Eq. (\ref{GW}) is obtained by considering that during preheating the
equation of state jumps from $\omega =0$ to an intermediate value close to $%
\omega =1/3$. Furthermore, $\omega $\ will reach $1/3$ just after preheating 
\cite{C8}\ which leads to $\left( a_{pre}/a_{re}\right) ^{1-1/4(1+\omega
)}=1 $. The current density fraction of radiation $\Omega _{r,0}=\rho
_{r,0}/\rho _{c,0}$ is given by $\Omega _{r,0}\simeq 9.1\times 10^{-5}$. $%
\Omega _{GW,0}h^{2}\propto 1/a_{0}^{4},$\ $g_{\ast }$\ and $g_{0}$\
satisfies $g_{\ast }/$\ $g_{0}\simeq 106.75/3.36\simeq 31$. The recent
analysis of 12.5-year of PTA published by the NANOGrav collaboration shows a
stochastic GWs behaviour. this analysis suggests a power-law type of the
strain of the GWs \cite{C9,C10,C11,C12} 
\begin{equation}
\Omega _{GW}(f)=\frac{2\pi ^{2}f_{yr}^{2}}{3H_{0}^{2}}A_{GWB}^{2}\left( 
\frac{f}{f_{yr}}\right) ^{\alpha }.
\end{equation}%
The NANOGrav measurements preferred ranges of parameters $A_{GWB}$ \cite{C12}
while according to \cite{D21} $\alpha =\left( 5-\gamma \right) \in \lbrack
-1.5,0.5]$, $f_{yr}$ is best estimated to $f_{yr}\simeq 3.1\times 10^{-8}$
and $H_{0}$\ is given by $H_{0}\equiv 100h~km/s/Mpc$.

\begin{figure*}[h]
\resizebox{1\textwidth}{!}{  \includegraphics{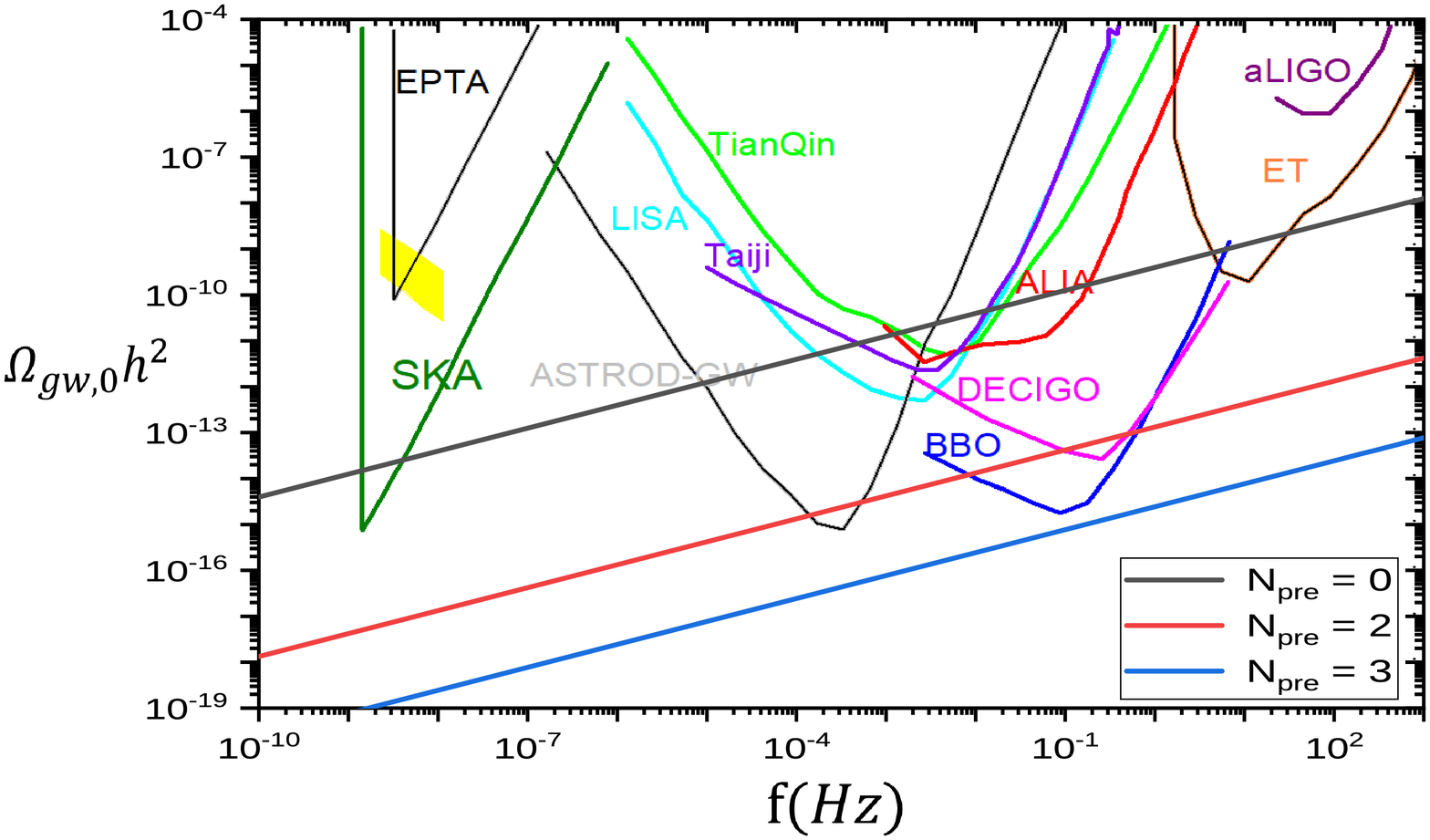}
}
\caption{The abundance of gravity wave energy density as function of the
present value of the frequency. We display, from \protect\cite%
{D15,D16,D17,D18,D19,D20}, the sensitivity curves of the Square Kilometer
Array (SKA), European Pulsar Timing Array (EPTA), Astrodynamical Space Test
of Relativity using Optical-GW detector (ASTROD-GW), Laser Interferometer
Space Antenna (LISA), Advanced Laser Interferometer Antenna (ALIA), Big Bang
Observer (BBO), Deci-hertz Interferometer GW Observatory (DECIGO), Advanced
LIGO (aLIGO) and in yellow NANOGrav 12.5 years experiments. We take $\protect%
\alpha =1/2$ and three different values of preheating duration, $N_{pre}=0$
(black), $2$ (red) and $3$ (blue).}
\label{fig:6}
\end{figure*}
In Fig. (\ref{fig:6}), we show the abundance of the energy density of
gravitational waves as a function of the present value of the frequency. We
also display some expected curves of GW experiments from \cite%
{D15,D16,D17,D18,D19,D20}. We numerically compute the energy spectrum of the
induced GWs and test the effect of the preheating duration on the GWs
spectrum. The GW spectrum increases the frequencies and spans over some
sensitivity curves for $N_{pre}\leq 2$. We conclude that our teleparallel
gravity model can explain the energy spectrum at high frequencies for $%
N_{pre}\leq 2$ such as ASTROD-GW, BBO and DECIGO. However, higher preheating
duration are not consistent with the expected GWs observatories.
Unfortunately, our teleparallel gravity model is far away to explain the
NANOGrav experiments. Moreover, the parameter $\delta $\ of our $f(T)$\
model can strongly affect the duration of preheating. According to our
previous results, the bound $\delta >3$\ predict higher values of $N_{pre}$
which can give inconsistant results with the GWs observations curves.
Furthermore, the case $\delta =3$\ provides compatible results with the blue
tilted energy spectrum of stochastic gravitational waves for higher
frequencies. The reason behind this is that our model of GWs spectrum is
derived taking into consideration the phase of preheating rather than
inflation alone. Here, we mention that the inflationary gravitational waves
are one of the most important stochastic gravitational wave background
sources that are used to probe the Universe and can provide possible
explanation for the recent NANOGrav results. However, our model predicts
higher frequencies because we considered subsequent phases following
inflation. \ 

\section{Conclusion}

\label{secVI}

In this paper, we have reviewed briefly the teleparallel gravity approach in
which we have related the slow roll and perturbatives parameters of the
inflationary paradigm. We have also introduced the parametric resonance in
the preheating era in which the production of the $\chi$ fields and the
growth of its fluctuations are mandatory for the production of the
primordial black holes and gravitational waves.

We have found that, for a chaotic inflation, a good consistency were
observed in relation with the $(r,n_s)$ plane at 1$\sigma$ for and 2$\sigma$
C.L. for some parameters of teleparallel gravity. We have also obtained that
the duration of the preheating is consistent with the $(r,n_s)$ plane at 1$%
\sigma$ for and 2$\sigma$ C.L..

Moreover, we have also calculated the abundance of primordial black holes
produced in the Teleparallel gravity model using the fraction of the energy
density and the variance of the density perturbations approach to study PBH
collapse. We have found that the primordial black holes overproduction can
be satisfied for specific values of the non-adiabatic curvature power
spectrum.

We have also obtained that the duration of the preheating is consistent with
the observational bound on $n_{s}$ and it must be less or equal to $2$
explain the present energy density of gravitational waves expected by
gravitational waves experiments.\newline
We have calculated the energy spectrum of gravitational waves numerically
and have shown that it presents an increasing behavior towards high
frequencies. We have shown that for the duration of the preheating equal to $%
2$, our model may explain the sensitivity curves expected by ASTEOD-GW, BBO
and DECIGO. On the other hand, instant transition of preheating i.e. $%
N_{pre}=0$, our model may explain the sensitivity curves of several
gravitational waves experiments, which gives the possibility for our chosen $%
f(T)$\ model to be examined by the energy spectrum of stochastic
gravitational waves observations.


\begin{thebibliography}{99}
\bibitem{will-2014} C. M. Will, The Confrontation between General Relativity
and Experiment, Living Rev. Rel. 17, 4 (2014) [1403.7377].

\bibitem{ligo-BH2016} LIGO Scientific, Virgo collaboration, Observation of
Gravitational Waves from a Binary Black Hole Merger, Phys. Rev. Lett. 116,
061102 (2016) [1602.03837].

\bibitem{ligo-GW2016} LIGO Scientific, Virgo collaboration, Tests of general
relativity with GW150914, Phys. Rev. Lett. 116, 221101 (2016) [1602.03841].

\bibitem{ligo2019} The LIGO Scientific Collaboration and the Virgo
Collaboration collaboration, Tests of general relativity with the binary
black hole signals from the ligo-virgo catalog gwtc-1, Phys. Rev. D 100,
104036 (2019).

\bibitem{SN-1999} Supernova Cosmology Project collaboration, Measurements of 
$\Omega$ and $\Lambda$ from 42 high redshift supernovae, Astrophys. J. 517,
565 (1999) [astro-ph/9812133].

\bibitem{SN-1998} Supernova Search Team collaboration, Observational
evidence from supernovae for an accelerating Universe and a cosmological
constant, Astron. J. 116, 1009 (1998) [astro-ph/9805201].

\bibitem{WMAP-2003} WMAP collaboration, First year Wilkinson Microwave
Anisotropy Probe (WMAP) observations: Determination of cosmological
parameters, Astrophys. J. Suppl. 148, 175 (2003) [astro-ph/0302209].

\bibitem{Scolnic-2018} D. M. Scolnic et al., The Complete Light-curve Sample
of Spectroscopically Confirmed SNe Ia from Pan-STARRS1 and Cosmological
Constraints from the Combined Pantheon Sample, Astrophys. J. 859, 101 (2018)
[1710.00845].

\bibitem{planck-2020} Planck collaboration, Planck 2018 results. VI.
Cosmological parameters, Astron. Astrophys. 641, A6 (2020) [1807.06209].

\bibitem{Guth-1918} A. H. Guth, The Inflationary Universe: A Possible
Solution to the Horizon and Flatness Problems, Phys. Rev. D 23, 347 56
(1981).

\bibitem{Linde-1982} A. D. Linde, The New Inflationary Universe Scenario: A
Possible Solution of the Horizon, Flatness, Homogeneity, Isotropy and
Primordial Monopole Problems, Phys. Lett. B 108, 389 93 (1982). 

\bibitem{A1} Linde, Andrei. "Inflationary cosmology." Lect. Notes Phys. 738,
1-54 (2008) [0705.0164].

\bibitem{A2} Gorbunov, Dmitry S., and Valery A. Rubakov. Introduction to the
theory of the early Universe: Cosmological perturbations and inflationary
theory. World Scientific, 2011.

\bibitem{A3} Lyth, David H., and Antonio Riotto. "Particle physics models of
inflation and the cosmological density perturbation." Physics Reports. 314,
1-146 (1999).

\bibitem{A4} Aghanim, N., Planck collaboration, Planck 2018 results. VI.
Cosmological parameters. 641, A6 (2020) [1807.06209]. 

\bibitem{Weinberg-1989} S. Weinberg, The cosmological constant problem, Rev.
Mod. Phys. 61 (1989) 1.

\bibitem{Malquarti-2003} M. Malquarti, E. J. Copeland and A. R. Liddle,
K-essence and the coincidence problem, Phys. Rev. D 68, 023512 (2003)
[astro-ph/0304277].

\bibitem{Verde-2019} L. Verde, T. Treu and A. Riess, Tensions between the
Early and the Late Universe, Nature Astron. 3, 891 (2019) [1907.10625].

\bibitem{Riess-2021} A. G. Riess, S. Casertano, W. Yuan, J. B. Bowers, L.
Macri, J. C. Zinn et al., Cosmic Distances Calibrated to 1%
Photometry of 75 Milky Way Cepheids Confirm Tension with $\Lambda$CDM,
Astrophys. J. Lett. 908, L6 (2021) [2012.08534].

\bibitem{Clifton-2012} T. Clifton, P. G. Ferreira, A. Padilla and C.
Skordis, Modified Gravity and Cosmology, Phys. Rept. 513, 1 (2012)
[1106.2476].

\bibitem{Kase-2019} R. Kase and S. Tsujikawa, Dark energy in Horndeski
theories after GW170817: A review, Int.J. Mod. Phys. D 28, 1942005 (2019)
[1809.08735].

\bibitem{Kobayashi-2019} T. Kobayashi, Horndeski theory and beyond: a
review, Rept. Prog. Phys. 82(8),086901 (2019) [1901.07183].

\bibitem{Bahamonde-2021} S. Bahamonde, K. F. Dialektopoulos, C.
Escamilla-Rivera, G. Farrugia, V. Gakis, M. Hendry et al., Teleparallel
Gravity: From Theory to Cosmology, [arXiv:2106.13793v3 [gr-qc]].

\bibitem{CANTATA-2021} CANTATA collaboration, Modified Gravity and
Cosmology: An Update by the CANTATA Network, [arXiv:2105.12582 [gr-qc]].

\bibitem{Felice2010} A. De Felice and S. Tsujikawa, Living Rev. Rel. 13, 3
(2010) [arXiv:1002.4928 [gr-qc]].

\bibitem{Nojiri2005} S. Nojiri and S. D. Odintsov, Phys. Lett. B 631, 1
(2005) [hep-th/0508049].

\bibitem{Hayashi-1967} K. Hayashi and T. Nakano, Prog. Theor. Phys. 38,
491--507 (1967).

\bibitem{Hayashi1982} K. Hayashi and T. Shirafuji, Phys. Rev. D 19, 3524
(1979); Addendum-ibid. 24, 3312 (1982).

\bibitem{Aldrovandi2013} R. Aldrovandi, J.G. Pereira, Teleparallel Gravity:
An In- troduction, Springer, Dordrecht, 2013.

\bibitem{Maluf2013} J. W. Maluf, Annalen Phys. 525, 339 (2013)
[arXiv:1303.3897 [gr-qc]]. 

\bibitem{I1} Cai, Y. F., Capozziello, S., De Laurentis, M., \& Saridakis, E.
N., f (T) teleparallel gravity and cosmology. Reports on Progress in
Physics, 79(10), 106901 (2016).

\bibitem{Li2011} M. Li, R. X. Miao and Y. G. Miao, JHEP 1107, 108 (2011)
[arXiv:1105.5934 [hep-th]].

\bibitem{Bamba2012} K. Bamba, R. Myrzakulov, S. Nojiri and S. D. Odintsov,
Phys. Rev. D 85, 104036 (2012) [arXiv:1202.4057 [gr-qc]].

\bibitem{Wu-2012} Y. P. Wu and C. Q. Geng, Phys. Rev. D 86, 104058 (2012)
[arXiv:1110.3099 [gr-qc]].

\bibitem{Ong2013} Y. C. Ong, K. Izumi, J. M. Nester and P. Chen, Phys. Rev.
D 88, 024019 (2013) [arXiv:1303.0993 [gr-qc]].

\bibitem{Otalora2013} G. Otalora, JCAP 1307, 044 (2013) [arXiv:1305.0474
[gr- qc]].

\bibitem{Bahamonde2015} S. Bahamonde, C. G. Boehmer and M. Wright, Phys.

Rev. D 92, no. 10, 104042 (2015) [arXiv:1508.05120 [gr- qc]].

\bibitem{Rezazadeh2016} K. Rezazadeh, A. Abdolmaleki and K. Karami, JHEP
1601, 131 (2016) [arXiv:1509.08769 [gr-qc]].

\bibitem{Farrugia2016} G. Farrugia and J. L. Said, Phys. Rev. D 94, no. 12,
124054 (2016) [arXiv:1701.00134 [gr-qc]].

\bibitem{Awad2017} A. M. Awad, S. Capozziello and G. G. L. Nashed, JHEP
1707, 136 (2017) [arXiv:1706.01773 [gr-qc]]. 

\bibitem{Rezazadeh-2017} K. Rezazadeh, A. Abdolmaleki and K. Karami,
Astrophys. J. 836, 228 (2017) [arXiv:1702.07877 [gr-qc]].

\bibitem{B3} Goodarzi, P., \& Sadjadi, H. M., Reheating in a modified
teleparallel model of inflation. The European Physical Journal C, 79 (3),
1-9 (2019).


\bibitem{Car2016} B. Carr, F. Kuhnel and M. Sandstad, Phys. Rev. D 94, no.8,
083504 (2016) doi:10.1103/PhysRevD.94.083504 [arXiv:1607.06077
[astro-ph.CO]].

\bibitem{Gaggero2017} D. Gaggero, G. Bertone, F. Calore, R. M. T. Connors,
M. Lovell, S. Markoff and E. Storm, Phys. Rev. Lett. 118, no.24, 241101
(2017) doi:10.1103/PhysRevLett.118.241101 [arXiv:1612.00457 [astro-ph.HE]].

\bibitem{Inomata2017} K. Inomata, M. Kawasaki, K. Mukaida, Y. Tada and T. T.
Yanagida, Phys. Rev. D 96, no.4, 043504 (2017)
doi:10.1103/PhysRevD.96.043504 [arXiv:1701.02544 [astro-ph.CO]].

\bibitem{Kovetz2017} E. D. Kovetz, Phys. Rev. Lett. 119, no.13, 131301
(2017) doi:10.1103/PhysRevLett.119.131301 [arXiv:1705.09182 [astro-ph.CO]].

\bibitem{Georg2017} J. Georg and S. Watson, JHEP 09, 138 (2017)
doi:10.1007/JHEP09(2017)138 [arXiv:1703.04825 [astro-ph.CO]] 

\bibitem{C1} Zhou, Z., Jiang, J., Cai, Y. F., Sasaki, M., \& Pi, S.,
Primordial black holes and gravitational waves from resonant amplification
during inflation. Physical Review D 102 (10), 103527 (2020).

\bibitem{C2} Carr, B., Dimopoulos, K., Owen, C., \& Tenkanen, T., Primordial
black hole formation during slow reheating after inflation. Physical Review
D 97 (12), 123535 (2018).

\bibitem{C3} Green, A. M., \& Malik, K. A., Primordial black hole production
due to preheating. Physical Review D 64 (2), 021301 (2001).


\bibitem{Belotsky2014} K. M. Belotsky, A. D. Dmitriev, E. A. Esipova, V. A.
Gani, A. V. Grobov, M. Y. Khlopov, A. A. Kirillov, S. G. Rubin and I. V.
Svadkovsky, Mod. Phys. Lett. A 29, no.37, 1440005 (2014)
doi:10.1142/S0217732314400057 [arXiv:1410.0203 [astro-ph.CO]].

\bibitem{Khlopov2010} M. Y. Khlopov, Res. Astron. Astrophys. 10, 495-528
(2010) doi:10.1088/1674-4527/10/6/001 

\bibitem{6} Lawrence Krauss, Scott Dodelson, and Stephan Meyer. Primordial
Gravitational Waves and Cosmology. Science 328, 989-992 (2010).

\bibitem{3} Lev Kofman, Andrei Linde and Alexei A. Starobinsky, "REHEATING
AFTER INFLATION", Phys.Rev.Lett.73, 3195-3198 (1994).

\bibitem{Khlebnikov} S. Yu. Khlebnikov and I. I. Tkachev, "Relic
gravitational waves produced after preheating", Phys. Rev. D 56 (2), 653
(1997).

\bibitem{2} Jing Liu, Zong-Kuan Guo, Rong-Gen Cai and Gary Shiu,
"Gravitational Waves from Oscillons with Cuspy Potentials", Phys. Rev. Lett.
120, 031301 (2018).

\bibitem{4} Richard Easther and Eugene A. Lim, "Stochastic Gravitational
Wave Production After Inflation", Journal of Cosmology and Astroparticle
Physics 2006(04), 010 (2006).

\bibitem{5} Richard Easther, John T. Giblin, Jr, and Eugene A.
Lim,"Gravitational Wave Production At The End Of Inflation", Physical Review
Letters 99 (22), 221301 (2007).

\bibitem{B7} Pereira, J. G., \& Obukhov, Y. N., Gauge structure of
teleparallel gravity. Universe 5(6), 139 (2019).

\bibitem{B7-1} Bahamonde, S., Dialektopoulos, K. F., Escamilla-Rivera, C.,
Farrugia, G., Gakis, V., Hendry, M., ... \& Di Valentino, E., Teleparallel
gravity: from theory to cosmology. arXiv preprint arXiv:2106.13793 (2021).

\bibitem{B1} Basilakos, S., Capozziello, S., De Laurentis, M.,
Paliathanasis, A., \& Tsamparlis, M., Noether symmetries and analytical
solutions in $f(T)$ cosmology: A complete study. Physical Review D 88 (10),
103526 (2013).

\bibitem{B2} Rezazadeh, K., Abdolmaleki, A., \& Karami, K., Logamediate
inflation in f (T) teleparallel gravity. The Astrophysical Journal 836 (2),
228 (2017).

\bibitem{D10} Creminelli, P., Nacir, D. L., Simonovi\'{c}, M., Trevisan, G.,
\& Zaldarriaga, M., $\phi^{2}$ or not $\phi^{2}$ : testing the simplest
inflationary potential. Physical Review Letters 112 (24), 241303 (2014).

\bibitem{D12} Green, A. M., \& Malik, K. A., Primordial black hole
production due to preheating. Physical Review D 64 (2), 021301 (2001).

\bibitem{D4} Harada, T., Yoo, C. M., \& Kohri, K., Threshold of primordial
black hole formation. Physical Review D 88 (8), 084051 (2013).

\bibitem{D11} Liddle, A. R., Lyth, D. H., Malik, K. A., \& Wands, D.,
Super-horizon perturbations and preheating. Physical Review D 61 (10),
103509 (2000).

\bibitem{B5} El Bourakadi, K., Ferricha-Alami, M., Filali, H., Sakhi, Z., \&
Bennai, M., Gravitational waves from preheating in Gauss--Bonnet inflation.
The European Physical Journal C 81 (12), 1-8 (2021).

\bibitem{B5-1} Kofman, L., Preheating after inflation. In COSMO-97 (pp.
312-321) (1998).

\bibitem{B5-2} Goodarzi, P., \& Sadjadi, H. M., Reheating in a modified
teleparallel model of inflation. The European Physical Journal C, 79(3), 1-9
(2019).

\bibitem{B5-3} Shtanov, Y., Traschen, J., \& Brandenberger, R., Universe
reheating after inflation. Physical Review D, 51(10), 5438 (1995).

\bibitem{B5-4} Mielczarek, J., Reheating temperature from the CMB. Physical
Review D, 83(2), 023502 (2011).

\bibitem{B5-5} Linde, A., Particle physics and inflationary cosmology (Vol.
5). CRC press (1990).

\bibitem{B6} El Bourakadi, K., Bousder, M., Sakhi, Z., \& Bennai, M.,
Preheating and reheating constraints in supersymmetric braneworld inflation.
The European Physical Journal Plus 136 (8), 1-19 (2021).

\bibitem{D3} Carr, B. J., The Primordial black hole mass spectrum,
Astrophysical Journal 201, 1-19 (1975).

\bibitem{D13} Sanchez, N., \& Zichichi, A., Current topics in
astrofundamental physics. Current topics in astrofundamental physics/edited
by N. Sanchez, (1997).

\bibitem{D14} Green, A. M., \& Liddle, A. R., Constraints on the density
perturbation spectrum from primordial black holes. Physical Review D 56
(10), 6166 (1997).

\bibitem{D8} Bhaumik, N., \& Jain, R. K., Primordial black holes dark matter
from inflection point models of inflation and the effects of reheating.
Journal of Cosmology and Astroparticle Physics 2020 (01), 037 (2020).

\bibitem{D7} Hertzberg, M. P., \& Yamada, M., Primordial black holes from
polynomial potentials in single field inflation. Physical Review D 97 (8),
083509 (2018).

\bibitem{C4} Wu, Y. P., \& Geng, C. Q., Matter density perturbations in
modified teleparallel theories. Journal of High Energy Physics (11), 1-17
(2012).

\bibitem{C5} Izumi, K., \& Ong, Y. C., Cosmological perturbation in f (T)
gravity revisited. Journal of Cosmology and Astroparticle Physics (06), 029
(2013).

\bibitem{C6} Cai, Y. F., Li, C., Saridakis, E. N., \& Xue, L. Q., f (T)
gravity after GW170817 and GRB170817A. Physical Review D 97 (10), 103513
(2018).

\bibitem{C7} J.F. Dufaux, A. Bergman, G. Felder, L. Kofman, J.P. Uzan,
Theory and numerics of gravitational waves from preheating after inflation.
Phys. Rev. D 76.12, 123517 (2007).

\bibitem{C8} Lozanov, K. D., \& Amin, M. A., Equation of state and duration
to radiation domination after inflation. Physical Review Letters 119(6),
061301 (2017).

\bibitem{C9} Kohri, K., \& Terada, T., Solar-mass primordial black holes
explain NANOGrav hint of gravitational waves. Physics Letters B 813, 136040
(2021).

\bibitem{C10} Vaskonen, V., \& Veerm\"{a}e, H., Did NANOGrav see a signal
from primordial black hole formation?. Physical Review Letters 126 (5),
051303 (2021).

\bibitem{C11} De Luca, V., Franciolini, G., \& Riotto, A., NANOGrav data
hints at primordial black holes as dark matter. Physical Review Letters 126
(4), 041303 (2021).

\bibitem{C12} Arzoumanian, Z., Baker, P. T., Blumer, H., B\'{e}csy, B.,
Brazier, A., Brook, P. R., ... \& NANOGrav Collaboration., The NANOGrav 12.5
yr data set: search for an isotropic stochastic gravitational-wave
background. The Astrophysical journal letters 905 (2), L34 (2020).

\bibitem{D21} Gao, T. J., \& Yang, X. Y., Double peaks of gravitational wave
spectrum induced from inflection point inflation. The European Physical
Journal C, 81 (6), 1-10 (2021).

\bibitem{D15} Drees, M., \& Xu, Y., Overshooting, critical Higgs inflation
and second order gravitational wave signatures. The European Physical
Journal C, 81 (2), 1-22 (2021).

\bibitem{D16} Amaro-Seoane, P., Audley, H., Babak, S., Baker, J., Barausse,
E., Bender, P., ... \& Zweifel, P., Laser interferometer space antenna.,
(2017). arXiv preprint arXiv:1702.00786.

\bibitem{D17} Ruan, W. H., Guo, Z. K., Cai, R. G., \& Zhang, Y. Z., Taiji
program: Gravitational-wave sources. International Journal of Modern Physics
A, 35 (17), 2050075(2020).

\bibitem{D18} Moore, C. J., Cole, R. H., \& Berry, C. P., Gravitational-wave
sensitivity curves. Classical and Quantum Gravity, 32 (1), 015014 (2014).

\bibitem{D19} Luo, J., Chen, L. S., Duan, H. Z., Gong, Y. G., Hu, S., Ji,
J., ... \& Zhou, Z. B., TianQin: a space-borne gravitational wave detector.
Classical and Quantum Gravity, 33 (3), 035010 (2016).

\bibitem{D20} Kuroda, K., Ni, W. T., \& Pan, W. P., Gravitational waves:
Classification, methods of detection, sensitivities and sources.
International Journal of Modern Physics D, 24 (14), 1530031 (2015).
\end{thebibliography}
\end{document}